# White organic light-emitting diodes with an ultra-thin premixed emitting layer

T. Jeon[1], B. Geffroy[1,2], D. Tondelier[1], Y. Bonnassieux[1], S. Forget[3, 5 a)], S. Chenais[3, 5], E. Ishow[4]

[1] *LPICM, CNRS UMR 7647, Ecole Polytechnique, 91128, Palaiseau, France*

[2] *CEA Saclay, DSM/IRAMIS/SPCSI/LCSI, 91191 GiF Sur Yvette, France*

[3] *Université Paris 13, Sorbonne Paris Cité, Laboratoire de Physique des Lasers, F-93430, Villetaneuse, France*

[4] *PPSM, CNRS UMR 8531, ENS Cachan, 94235 Cachan, France*

[5] *CNRS, UMR 7538, LPL, F-93430, Villetaneuse, France*

a) *Corresponding author : sebastien.forget@univ-paris13.fr*

## ABSTRACT:

We described an approach to achieve fine color control of fluorescent White Organic Light Emitting Diodes (OLED), based on an Ultra-thin Premixed emitting Layer (UPL). The UPL consists of a mixture of two dyes (red-emitting 4-di(4'-*tert*-butylbiphenyl-4-yl)amino-4'-dicyanovinylbenzene or *fvin* and green-emitting 4-di(4'-*tert*-butylbiphenyl-4-yl)aminobenzaldehyde or *fcho*) premixed in a single evaporation cell: since these two molecules have comparable structures and similar melting temperatures, a blend can be evaporated, giving rise to thin films of identical and reproducible composition compared to those of the pre-mixture. The principle of fine color tuning is demonstrated by evaporating a 1-nm-thick layer of this blend within the hole-transport layer (4,4'-bis[N-(1-naphtyl)-N-

phenylamino]biphenyl (α-NPB)) of a standard fluorescent OLED structure. Upon playing on the position of the UPL inside the hole-transport layer, as well as on the premix composition, two independent parameters are available to finely control the emitted color. Combined with blue emission from the heterojunction, white light with Commission Internationale de l'Eclairage (CIE) 1931 color coordinates (0.34, 0.34) was obtained, with excellent color stability with the injected current. The spectrum reveals that the *fcho* material does not emit light due to efficient energy transfer to the red-emitting *fvin* compound but plays the role of a host matrix for *fvin*, allowing for a very precise adjustment of the red dopant amount in the device.

## 1. Introduction

Organic Light-Emitting Diodes (OLED) have made important progress during the last two decades particularly for applications in flat panel displays. The interest for White OLED (WOLED) has recently been pushed forward by the general lighting market; they appear as useful complements and even competitors of well-established fluorescent tubes and inorganic LEDs thanks to their unique features (lightweight "light sheets" emitting over a sizeable area, low cost, glare-free emission, deposition over potentially flexible/transparent substrates[1]) going along with high efficiencies and a low environmental cost in terms of manufacturing, operation and disposal compared to existing technologies[2].

OLED emitting white light were first reported by J. Kido et al. 15 years ago [3]. Now, WOLED are able to outperform incandescent light bulbs and even fluorescent tubes in terms of luminous efficiencies[4]. A good white light emitter should be defined by color coordinates close to the Planckian locus around the equi-energy white point (x = 0.33, y = 0.33 in the Commission International de l'Eclairage (CIE) 1931 diagram), with a Color Rendering Index (CRI) close to 100. Moreover these features should be independent on the injected current over a large span to allow for luminance adjustment. They should be stable during the device lifetime (no differential aging of the luminophores), and at last the OLED manufacturing process should allow for a high reproducibility and a precise tuning of the color. The color stability issue is prominent for all OLED devices but is of particular importance for WOLED as only a shift of more than 0.005-0.01 for the x- and the y- values of CIE coordinates, respectively, results in an appreciable visible shift in the rendered color [5]. This correspond to a four-step MacAdam ellipse, which is a typical measure of an acceptable color shift (even if the lower sensitivity of the human eye in this part of the colorimetric diagram make this definition maybe rather stringent for white emitters). Although the basic principle to obtain white light is always to mix either two or three emitters with complementary colors[6-8], three different global approaches can be used to achieve this goal, each with its advantages and drawbacks [2]: the lumiphores can be blended in a single layer [9-11], separated in different layers in the same OLED[12, 13], or contained in several formally independent devices each of which emits light of a different color [14], the latter being however excessively complex and costly compared to the two other approaches. For those first two categories, however, several technical issues have to be addressed. The most critical issue in producing white light is the accurate and reproducible control of the emitted color, as well as the absence of color shift upon voltage modification. In this context, there is a technical issue originating from the lowest-energy dopant (the red emitter in a Blue+Green+Red combination, or the yellow dopant in a Blue+Yellow mix) because it causes emission quenching from higher-energy lying lumiphores due to energy transfer. Moreover it has to be incorporated in very small amounts compared to the other lumiphores or hosts. In a single blend, this is because Förster resonant energy transfer is extremely efficient towards the lowest-gap material. In more complex multilayer devices, even though energy transfer is also often exploited to play on the color, low doping rates in host matrices have to be used in any case, because concentration quenching is strong with red emitters, often characterized by long π-conjugated backbones and prompt to strong π-stacking and/or dipole-dipole interactions.

The most widely-used technique to achieve the required very low doping rates is co-evaporation; low doping ratios are obtained upon setting large temperature differences between the host and the guest evaporation boats; in practice batch-to-batch reproducibility is a problem and co-evaporation becomes impracticable for doping rates below 0.3% [15].

A few solutions have been proposed to overcome this limitation: the first one consists in using ultra-thin layers of *pure* dopants [7, 16, 17], whose position with respect to a recombination zone defines the color in a very precise fashion through the interplay of Förster energy transfer and exciton diffusion [18, 19]. This technique is limited to ultrathin "layers" (of the order of 1 nm, which is also the typical size of an emitter, so that it may be thought of as a "localized doping" or a "delta-doping" scenario [16]) and is very sensitive to concentration quenching. The emission spectrum does not generally depend on the applied voltage in this case as the recombination zone remains strongly localized [16].

A different approach consists in using single white-emitting layers from different compounds "premixed" in a single evaporation boat. Jou *et al.* used a mixture of host and dopant [20] and obtained high color reproducibility with however a strong chromaticity dependence on the deposition sequence, attributed to the very different melting points between the guest and the host molecules. Similar conclusions can be drawn from the work of Steinbacher *et al.* [15] who realized a yellow OLED by premixing two dopants in a single cell. Shao *et al.* [21] used a mixture of NPD with various fluorescent dopants melted together as an organic solid solution formed by a rather complex high-temperature and high-pressure fusion process. Using a premixed blend offers a much higher process simplicity and reproducibility; however dopants should have similar evaporation rates to eliminate the time-varying composition of the boat during the evaporation process.

In this study, we propose to merge both advantages of the ultrathin layer approach (low current-dependence of color coordinates, fine control of chromaticity) with those of the premixed blend (reproducibility and fine control of very low doping rates) with fluorescent lumiphores that are weakly sensitive to quenching issues. In this way, we elaborated WOLED made of an ultra-thin layer of two mixed dopants, both of which having similar chemical structure and being able to be premixed with high homogeneity and evaporated with similar rates. We have then two "knobs" that we can "turn" to adjust the color, namely the thin layer position with respect to the recombination zone, and the premix composition. The Ultrathin Premixed Layer (UPL) consists of a mixture of two dipolar starbust triphenylamine molecules (4-di(4'-*tert*-butylbiphenyl-4-yl)aminobenzaldehyde and 4-di(4'-*tert*-butylbiphenyl-4-yl)amino-4'-dicyanovinylbenzene respectively dubbed *fcho* and *fvin* in the following), presenting very similar structures and thermal properties, and emitting in the green and red region respectively. After evaporation, the resulting UPL film composition relates to that of the initial powder mixture, which in turn controls the OLED emission properties. Pure white light has been obtained for a *fcho:fvin* ratio of 1:2 with 1931 CIE color coordinates (0.34, 0.34).

## 2. Experimental procedure

The device structure used in this study is shown in Fig. 1(a). All organic layers were evaporated on cleaned Indium Tin Oxide (ITO) coated glass substrates (~20 $\Omega$/sq) without breaking the vacuum (~$10^{-7}$ mbar or $10^{-5}$ Pa) at a deposition rate of 0.2 nm/s. The layer thicknesses were monitored with an in-situ quartz crystal. The active area of the devices defined by the aluminum cathode is 0.3 cm². Copper (II) pthalocyanine (CuPc), 4,4'-bis[N-

(1-naphtyl)-N-phenylamino]biphenyl (α-NPB), aluminum tris(8-quinolinolato) (Alq$_3$) were used as a hole injection layer, hole transport and electron transport layers respectively. (4,4'-bis(2,2'-diphenyl vinylene)-1,1'-biphenyl) (DPVBi) was employed as a blue emissive host material. The thickness of each layer (see figure 1a) was defined so that the expected recombination zone location corresponds to a maximum of the optical field for an emission in the blue range of the spectrum. In this device structure, the recombination zone is preferably located at the α-NPB/DPVBi interface due to the charge confinement imposed by the energetic barrier levels of these two materials ($\Delta E_{LUMO} \sim 0.4$ eV and $\Delta E_{HOMO} \sim 0.5$ eV) [7,13,17]. On the cathode side, lithium fluoride (LiF) was used to improve electron injection. The 1-nm-thick UPL film was evaporated at a distance $d$ from the α-NPB/DPVBi interface (by convention d<0 is for the UPL inside α-NPB and d>0 for the UPL in DPVBi). The layer thicknesses are controlled with a quartz balance with an accuracy of 10%. The UPL is a mixture of the two molecules (*fcho* and *fvin*) depicted in Fig. 1(b) and whose synthesis, thermal and spectroscopic characteristics have been reported in an earlier work [22]. They were directly mixed as powders in the same evaporation boat and their weight ratio was adjusted before evaporation under vacuum. Four mixture ratios *fcho:fvin* of 1:0; 1:2; 2:1 and 0:1 respectively were investigated. The main properties of *fcho* and *fvin* can be summarized as follows. First, these starburst triphenylamine molecules have similar structures except for their electron-withdrawing groups (EWG), and form time-stable glassy materials with evaporation temperatures around 90°C. Therefore, we used *fcho* and *fvin* as solids in the same evaporation boat without further dissolution treatments. Second, the nature of the EWG modifies the energy of the radiative intramolecular charge transfer excited state, which causes tuning of the emission color. The Highest Occupied Molecular Orbital (HOMO) and Lowest Unoccupied Molecular Orbital (LUMO) energy levels were -5.3 eV and -1.6 eV respectively for *fcho* and -5.4 eV and -2.4 eV respectively for *fvin*. The HOMO energy levels were obtained by cyclic voltammetry and the LUMO energy levels were calculated from the HOMO levels and optical band gaps measured from the onset of absorption spectra [23]. The similarity of the HOMO energy levels stems from the localization of the electronic density mainly on the triphenylamino core. In the opposite, the LUMO levels are electronically spread on the electron-withdrawing units, hence stronger stabilization of the charge transfer state is observed for *fvin* where the dicyanovinylene group exerts stronger electron-deficiency than the aldehyde group present in *fcho*. Third, the presence of a twisted triphenylamino core substituted by terminal bulky *tert*-butyl groups prevents the molecules from interacting with each other and creating deleterious radiationless aggregates, therefore leading to a reduced probability of self-quenching [24]. Finally the large Stokes shifts (up to 6050 cm$^{-1}$ and 4850 cm$^{-1}$ for solid films of *fcho* and *fvin* respectively) highly reduce the risks of emission re-absorption and self-quenching through energy transfer.

The UPL films were also deposited on glass substrates to record their UV/Visible absorption spectra as well as their photoluminescence (PL) properties. For PL measurements, 50-nm-thick *fcho* and *fvin* thin films were excited with a Continuous-wave blue laser at 405 nm. Electroluminescence spectra and chromaticity coordinates of OLED devices were measured by a PR655 SpectraScan spectroradiometer at a constant current density of 30 mA/cm$^2$. All the measurements were carried out under dark conditions and ambient atmosphere.

## 3. Results and discussion

### 3.1. Validity of the UPL concept

First of all, it is significant to verify the validity of the UPL concept, namely the proper composition of the films. Fig.2 shows the normalized UV/Visible absorption spectrum of the UPL films with four different mixture ratios. A first absorption band, centered around 335 nm, was observed for both compounds, and ascribed to π−π* transitions located on the (4-*tert*- butylbiphenyl) amino unit. A second band peaking at 375 and 500 nm for *fcho* and *fvin* respectively, characterized the radiative amino-EWG charge transfer [22]. Fig.2 shows that the absorption spectra of evaporated films match quite well the initial mixture, proving the validity of the UPL concept: firstly all bands are only from *fcho* and/or *fvin* with no formation of unexpected new molecular aggregates, and secondly the initial mix composition is preserved in the film, the visible band intensity varying according to the relative concentration ratio between both materials.

The PL properties of UPL films are shown in Fig.3. The PL spectra exhibit maxima around 490 nm and 620 nm for pure *fcho* and *fvin* respectively, in accordance with the spectroscopic characteristics of each molecule [22]. The PL spectra of the 1:2 and 2:1 mixtures are almost identical to the *fvin* PL spectrum, showing that Förster Resonant Energy transfer (FRET) from the green-emitting *fcho* to the red *fvin* is almost complete in this case.

### 3.2. A first "knob" to tune the color: the position of the UPL layer

In a first experiment, we evaporated a thin 1-nm-thick UPL of fixed composition (*fcho:fvin* = 1:2) at various distances $d$ from the recombination zone. As shown in figure 4, and in accordance with previous results [7], the spectrum is composed of two broad peaks, centered around 460 and 600 nm respectively, whose relative weights are highly modulated depending on the distance $d$ : when $d$ is large compared to the singlet exciton diffusion length and the Förster radius, which are both here of the order of a few nm, blue emission from the exciton recombination zone prevails, as very few holes (for d>0) or electrons (for d<0) are captured by the UPL. As NPB and DPVBi have similar optical spectra and electronic gaps, the emission from these two compounds is indistinguishable and is supposed to occur with comparable probabilities [7]. The values of exciton diffusion lengths in NPB (4.9 nm) and

DPVBi (8.7 nm) given in [7] account correctly for the more abrupt change in color observed in the NPB side.

### 3.2. A second "knob" to tune the color: the composition of the UPL layer

Now we fix the distance $d$ to 3 nm in the NPB side (d=-3 nm following our notations), which corresponds to the typical order of magnitude for both Förster radius and singlet exciton diffusion length, that is the region where the chromaticity is expected to vary the most with the ultrathin layer composition. Fig.5 shows electroluminescence spectra for different UPL film compositions, showing indeed a large spectral variation. The UPL-free OLED emits blue light peaking at 460 nm. In the case of a pure *fcho* layer, a single blue light spectrum was observed, only slightly red shifted compared to that of the α-NPB/DPVBi interface (with a maximum at 468 nm) with similar shapes (Fig.5.) This blue peak is likely to result from the combination of the light emitted by the excitons formed at the α-NPB/DPVBi interface and emitted by *fcho* excitons. When the UPL is made of pure *fvin*, the OLED emits orange-red light at 580 nm with negligible blue emission. The electroluminescence (EL) spectrum of *fvin* appears to be blue-shifted compared to the PL spectrum: this shift is due firstly to the optical field in the OLED which is maximal for blue wavelengths at the heterostructure and secondly to polarization effects, as discussed in [24], as NPB is less polar than *fvin* molecules leading to a lower Stokes shift.

For intermediate composition mixtures, a spectrum with a variable balance of blue and yellow/red emission is observed. For a *fcho:fvin* ratio of 1:2, white light with 1931 CIE color coordinates (0.34, 0.34) is obtained, with a CRI of 72, and a 1.2% external quantum efficiency. The (x,y) chromaticity coordinates are remarkably stable over a large range of luminance (from 600 to 1200 cd/m²) as shown in fig. 7a. For lower current densities, the (x,y) coordinates both increase but interestingly, as it appears in fig. 7b, they remain at equal distance from the Planckian locus, meaning that the emitted color is close to a "warm white" without visible green or pink hue. Another way of representing this color variation is to draw the evolution of the CIE coordinates in the CIE diagram and to compare them with the MacAdam ellipses of various steps (fig. 8). It appears that for a luminance greater than 600 cd/m² , the color of the emitted light stays within a MacAdam ellipse of step 1, meaning that the color variation for those luminances cannot be distinguished by the human eye. As color variation within a step 4 MacAdam ellipse is commonly admitted to be acceptable in the lighting area, we can stand that our OLED are color-stable down to around 250 cd/m² . Except for low luminances, the reported color shift is therefore not problematic at all in a domestic lighting source; it could even be chased in some cases as the spectrum dependence with the injected current mimics the behavior of an incandescent light bulb. The current-voltage and luminance-voltage characteristics are shown in Fig. 9a. The WOLED device shows a good diode behavior with a turn-on voltage (defined when luminance exceeds 0.1 cd/m²) around 4.5 V. As shown in Fig. 9b, the current efficiency at 2.3 cd/A is remarkably constant as a function of the current density. This very good stability is expected with stable fluorescent emitters, which validates the excellent stability of the molecular materials in the UPL layer.

### 3.3. Interpretation of the results

A picture of the main mechanisms at work is featured on fig. 10. As it is clear from fig. 3 and 5, the blue part in the EL spectra of fig. 5 are only due to excitons produced at the NPB/DPVBi interface, and not to *fcho* since the latter fully transfer their excitation to *fvin* molecules. These interface excitons may transfer their energy through a direct Förster energy transfer to fvin (process A), or to *fcho* first (process B) followed by a resonant transfer from *fcho* to *fvin* (process C). Note that *fcho* excitons can also be excited by singlet exciton diffusion from the interface. Holes are expected to easily pass through the *fcho* and *fvin* layer because the HOMO levels of *fvin*, *fcho* and α-NPB are quite similar.

To be more quantitative, we calculated the Förster energy transfer rate constant $k_{ET,i}$ given by [25]:

$$k_{ET,i} = \frac{9000\ln 10}{128\pi^5} \frac{\kappa^2 \phi_{host} J}{N_A n^4} \frac{1}{\tau R_i^6} = \frac{R_0^6}{\tau R_i^6} \qquad (1)$$

where $\kappa^2$ is an orientation factor (0.476 for random orientations of donor and acceptor dipole moments in the solid state), $\phi_{host}$ is the fluorescence quantum yield of the energy donor in the absence of the energy acceptor, $N_A$ is the Avogadro's constant, n is the refractive index of the host system and τ is the donor excited lifetime in the absence of the acceptor. J is the overlap integral between the donor fluorescence spectrum and the acceptor absorption spectrum. $R_i$ features the distance separating an isolated donor-acceptor pair and $R_0$ is the Förster radius for the respective guest-host system corresponding to the distance where fluorescence emission and resonant energy transfer are equiprobable deactivation processes. $\phi_{host}$ of DPVBi and *fcho* have been valued to be about 0.45 [26] and 0.81 [22] respectively. Refractive indices were assumed to be similar (n~1.7) [27]. It is worthwhile to quantitatively compare the Förster energy transfer rates for processes A and C of Fig.9. The Förster radius for the *fcho* to *fvin* energy transfer process is as high as 4.1 nm, due to considerable spectral overlap between *fcho* emission and *fvin* absorption [22]. This value is much higher than the spatial separation between *fcho* and *fvin* molecules inside the UPL, explaining why the energy transfer from *fcho* to *fvin* is almost complete, as observed experimentally under optical excitation in fig. 3.

Regarding the energy transfer between the interface excitons and *fvin*, we calculate $R_0=3.9$ nm, which is also higher than the 3 nm distance between the NPB/DPVBi interface and the UPL, leading to a very efficient Förster transfer. A contrario, the Förster radius between interface excitons and *fcho* molecules in the UPL is only 2.7 nm because of the weak spectral overlap between emission of the donor and absorption of the acceptor. This value is lower than the 3nm interface-UPL distance and leads to an incomplete Förster energy transfer. Nevertheless, this transfer is still consequent since the excitons diffuse also through the α-NPB layer and come in the vicinity of the *fcho* layer.

Although there is no distinguishable direct green emission of *fcho* excitons, they play an important role as illustrated in fig. 6. Indeed, the white OLED (with *fcho:fvin* = 1:2) is almost two times more efficient in terms of external quantum efficiency than the pure *fvin* OLED at the same current density. It can also be seen from fig. 5 that the white OLED emits more orange light in a configuration where there are less *fvin* molecules being evaporated. This suggests that *fcho* herein plays the role of a host for *fvin* molecules, helping *fvin* emitters to stay apart from each other (reduction of concentration quenching) and assisting them by efficiently releasing their energy through FRET. Although we realized "nondoped" OLED in the sense that a pure thin film of mixed dopants was evaporated, it has to be reminded that the 1-nm-thick UPL has a thickness comparable with molecular dimensions, meaning that the UPL realizes a "localized doping" or "delta-doping", in which however in-plane interactions between randomly-orientated emitters can affect the quantum efficiency. The UPL approach is then efficient in creating a local host/guest planar system, where the adjunction of *fcho* to the red *fvin* fluorophore helps reducing quenching and enhances the efficiency.

**4. Conclusion**

In conclusion, White OLED were realized with a concept of ultrathin premixed layer, in which two isostructural materials of comparable melting temperatures and evaporation rates were premixed and evaporated in a thin nanometric layer at a variable distance from a main recombination zone. Fine color tuning was demonstrated, by playing on two parameters: the first one is the distance separating the UPL from the exciton recombination zone, the second one is the UPL composition. With a *fcho*(green):*fvin*(red) ratio of 1:2 UPL evaporated at 3 nm of the NPP/DPBVi interface in the NPB side, we obtained white OLED with CIE (x,y)=(0.34,0.34) and good color stability for luminance >600 cd/m². Below this value, a color shift was observed but the color remained a warm white lying at equal distance from the Planckian locus. The spectra were proved to be governed by the red dopant (fvin) emission and blue excitons generated at the heterostructure, the green-emitting fcho molecules playing the role of a host in a host/guest system, i.e. enhancing the efficiency through reduction of concentration quenching and energy transfer. The UPL is a powerful concept that can be applied in all cases where a fine control of emission characteristics is highly sought and low doping rates are difficult to achieve experimentally.

**Acknowledgments**

Financial support by the CNRS Materials Program (MaProSu 2010- project M2EM) is gratefully acknowledged.

**FIGURE CAPTIONS**

Fig.1. (a) OLED device structure and (b) structure of the dipolar emissive triphenylamine derivatives.

Fig.2. UV/Visible absorption measurement of emissive films made of four different fcho: fvin mixtures evaporated on glass substrates.

Fig.3. Photoluminescence of emissive films with four different ratios fcho: fvin.

Fig.4 : Spectra recorded for OLED with an Ultrathin Premixed Layer of fixed composition (*fcho:fvin* = 1:2) placed at various positions *d* from the NPB/DPVBi exciton recombination surface (by convention : d<0 in NPB, d>0 in DPVBi)

Fig.5. Electroluminescence spectra of UPL films inserted in α-NPB.

Fig.6. Variation of the External Quantum Efficiency for 4 UPL compositions (fvin ratio into fcho of 0, 1/3, 2/3 and 1) in OLED with the UPL placed at d = - 3 nm. The fcho:fvin 1:2 composition corresponds to the white OLED.

Fig.7. (left) Variation of the CIE 1931 *x* and *y* coordinates for the white OLED (*fcho:fvin* ratio 1:2, d = -3 nm) as a function of the total emitted luminance. (right) : Position of (x,y) coordinates with respect to the Planckian locus.

Fig.8 : evolution of the CIE coordinates with the luminance (value written near each point, in cd/m²) and comparison with the MacAdam ellipses of various steps

Fig.9. WOLED performance (UPL – fcho: fvin = 1: 2). (a) J-V-L characteristics, (b) current efficiency versus current density.

Fig.10. Schematic emission mechanisms of OLED within the UPL film. Process A features exciton diffusion and energy transfer from the α-NPB/DPVBi interface to fvin (*favorable*). Process B corresponds to exciton diffusion and energy transfer from the α-NPB/DPVBi interface to fcho (*unfavorable*). Process C represents energy transfer from fcho to fvin. Processes A and C are in competition.

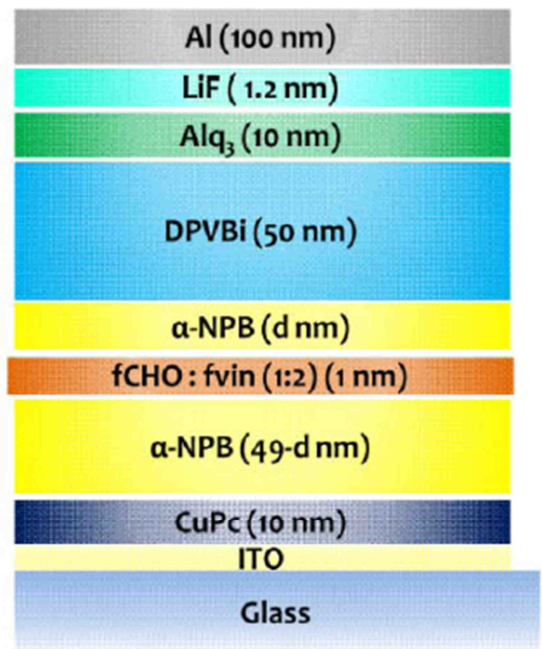

(a)

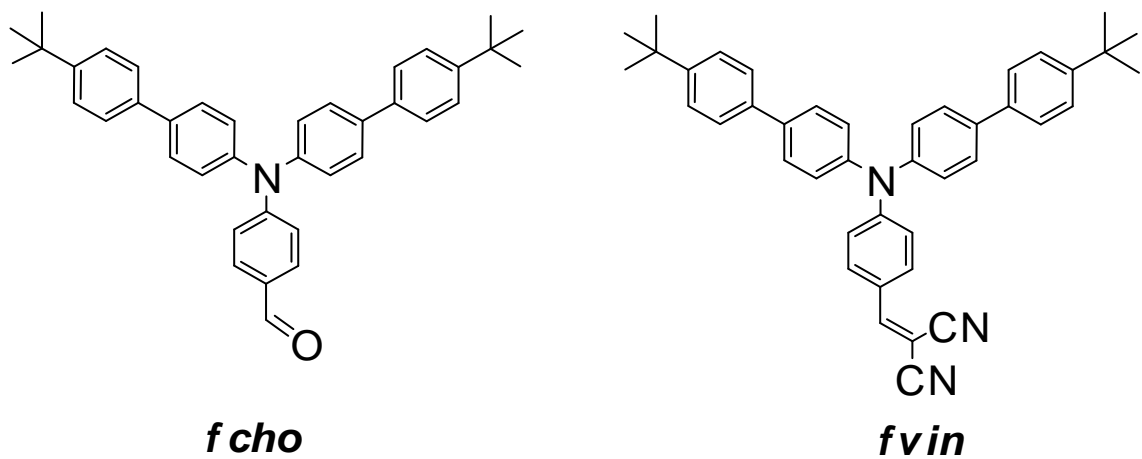

*f cho*          *f v in*

(b)

Fig.1.

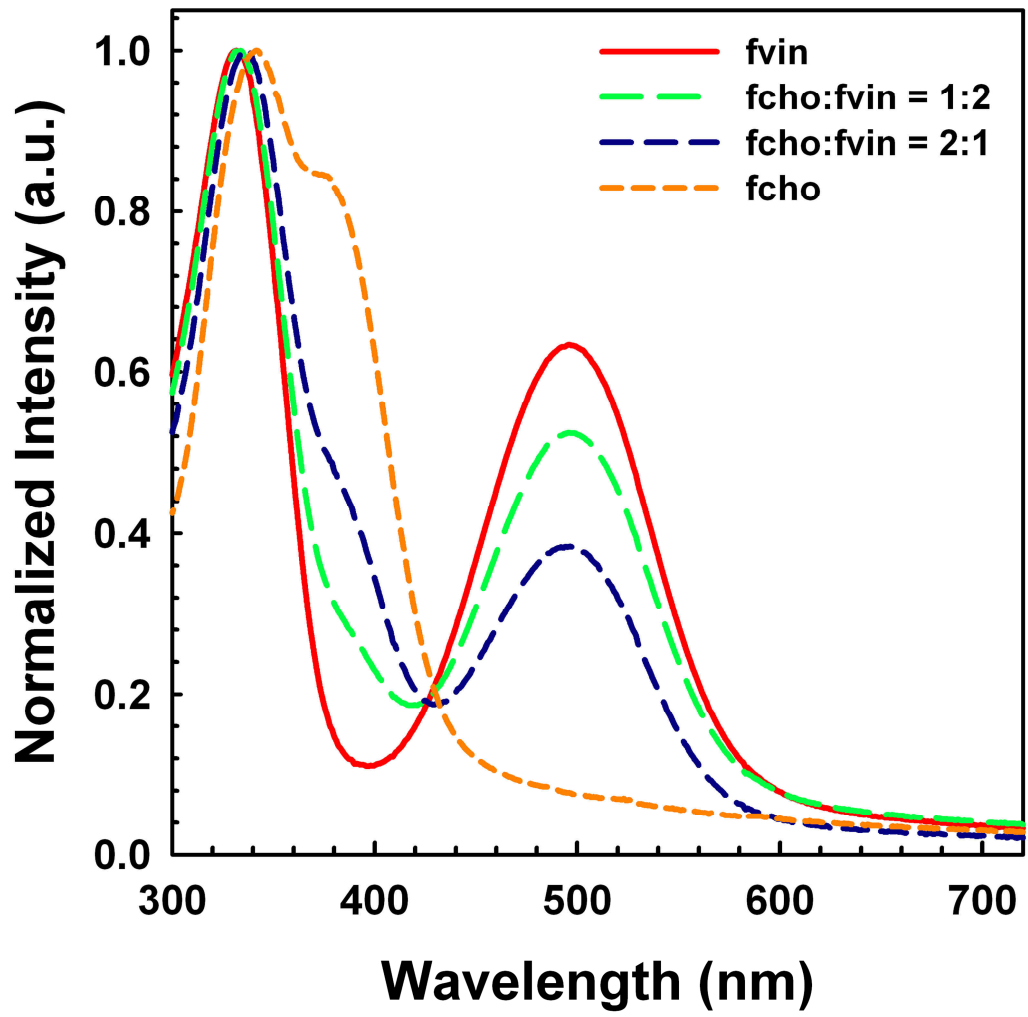

Fig.2.

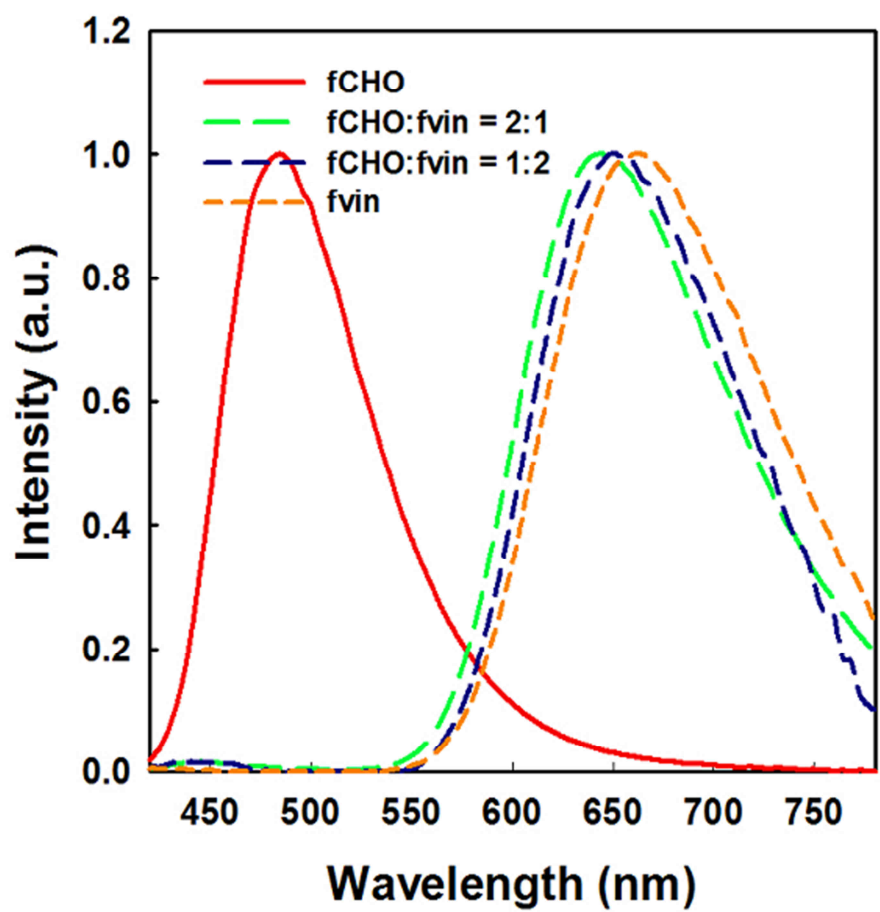

Fig.3.

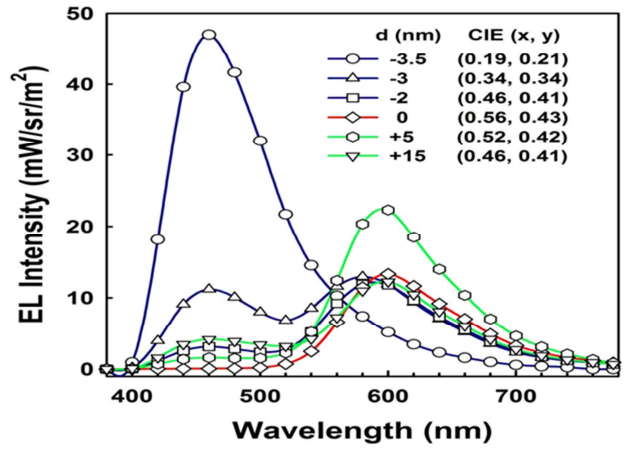

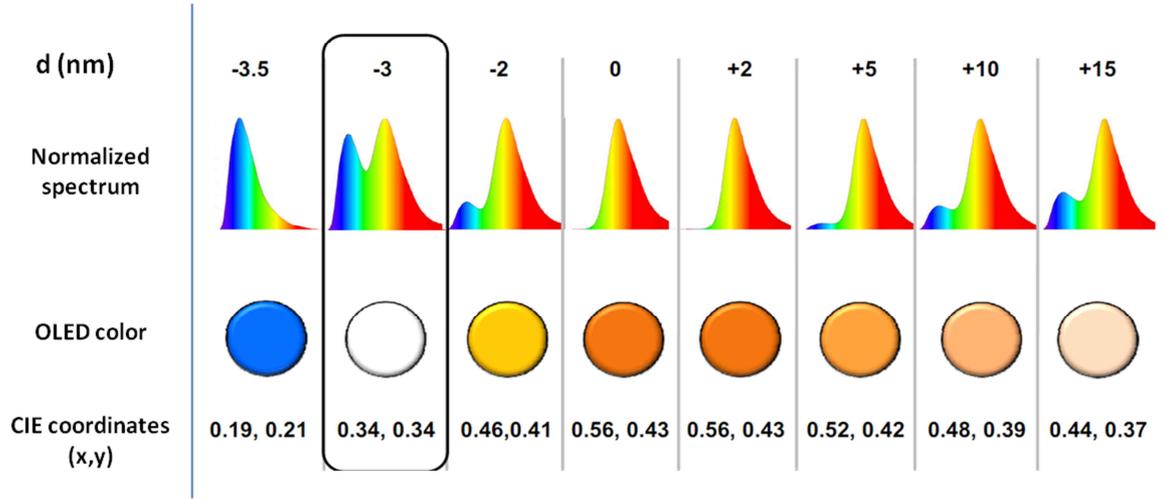

Fig.4 :

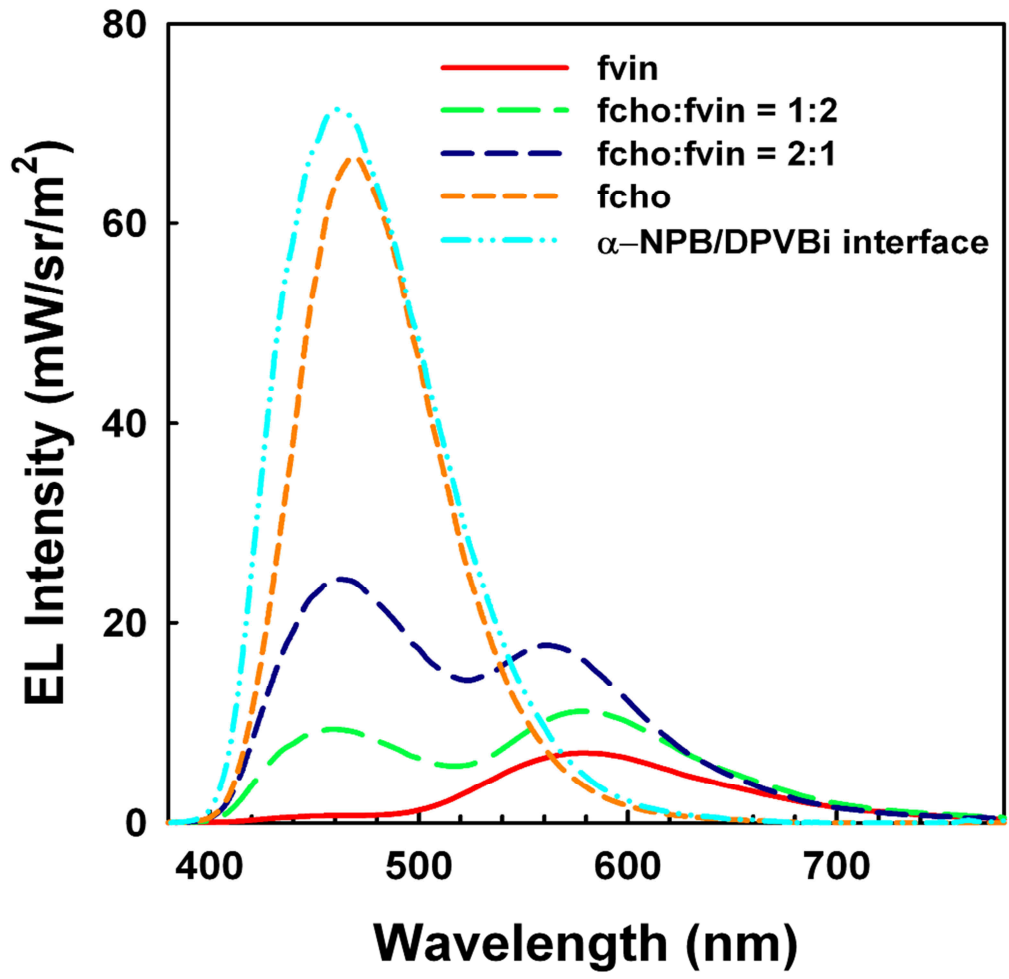

Fig.5.

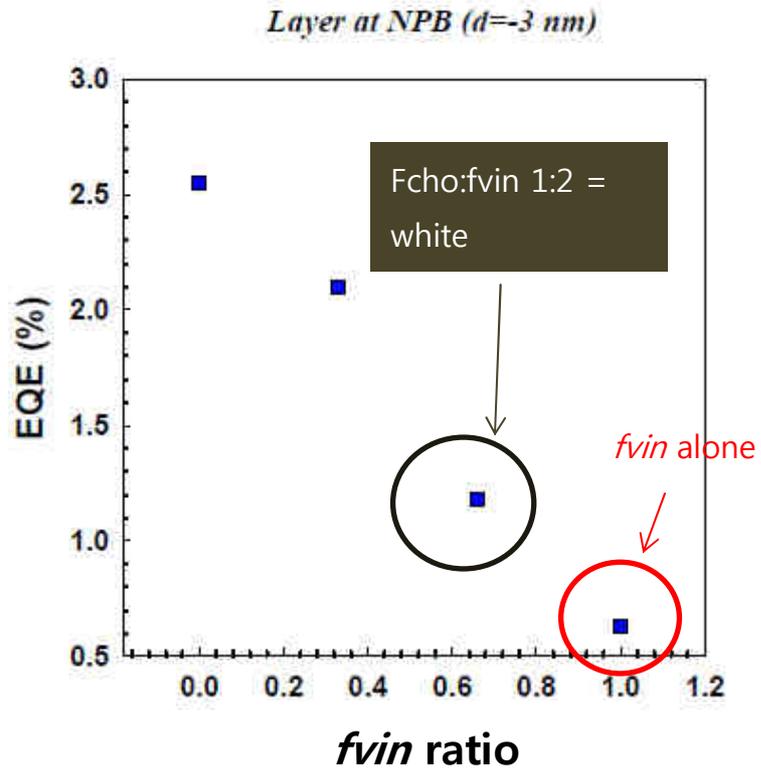

Fig.6.

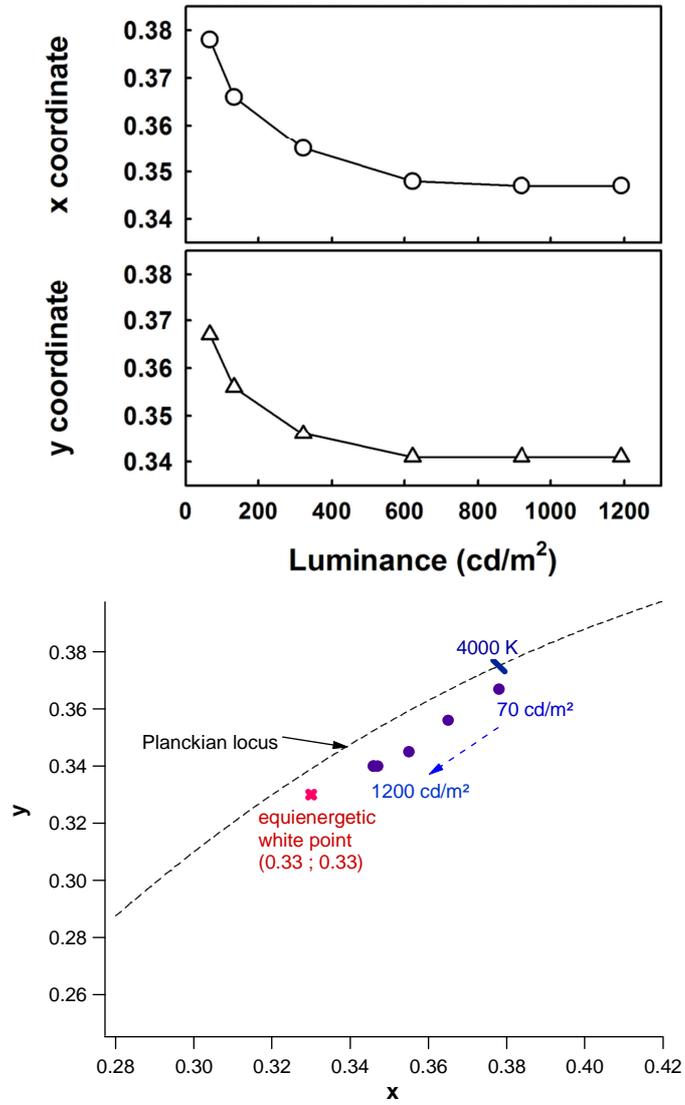

Fig.7.

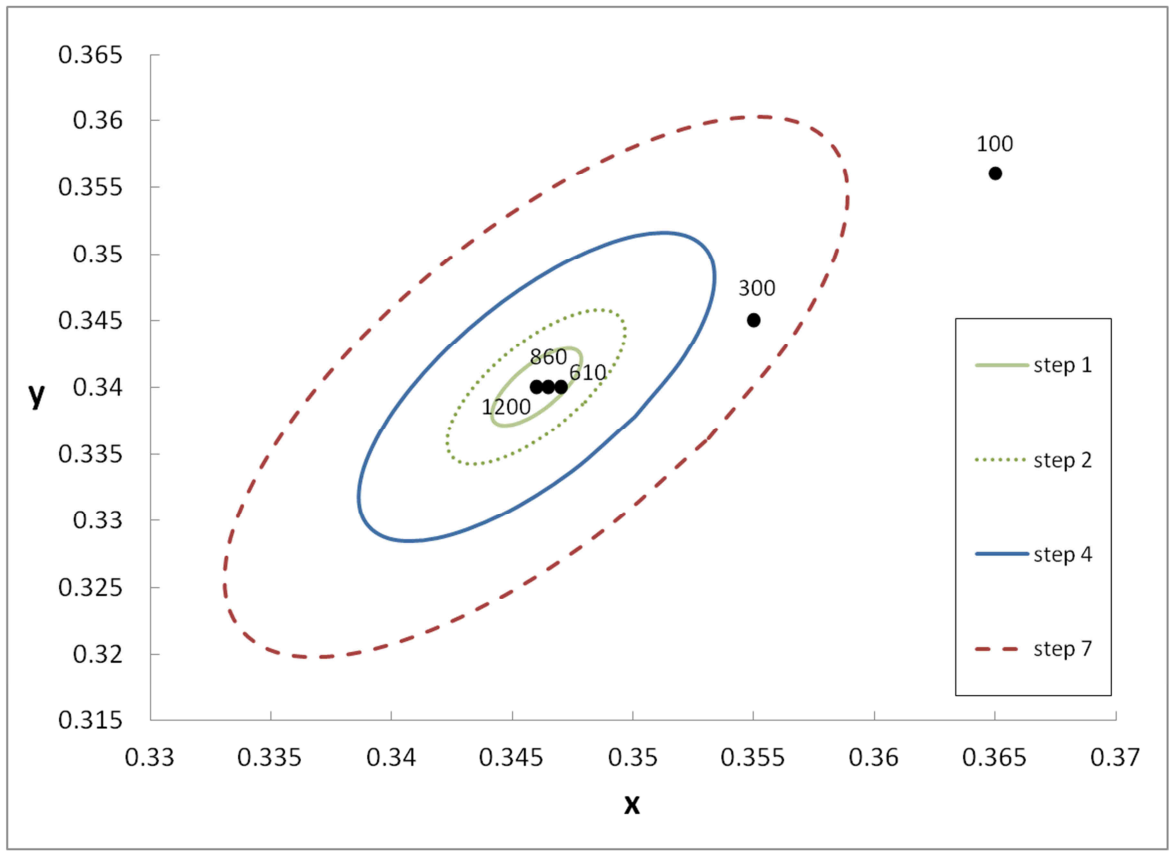

Fig.8.

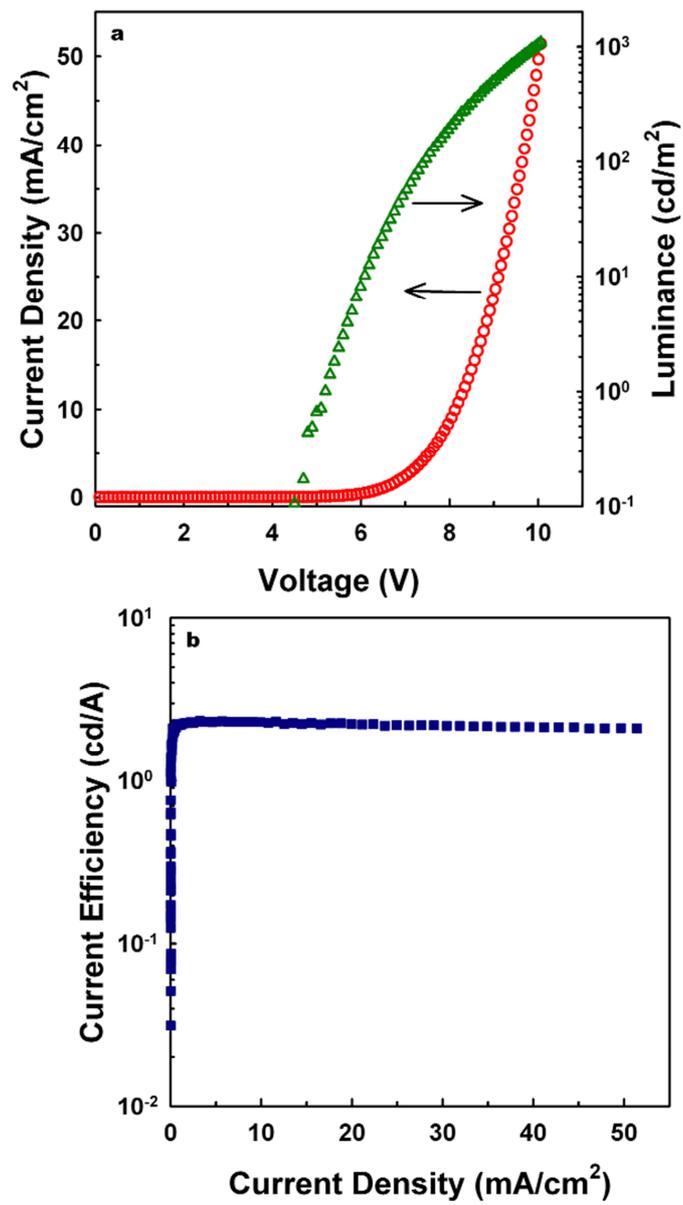

Fig.9

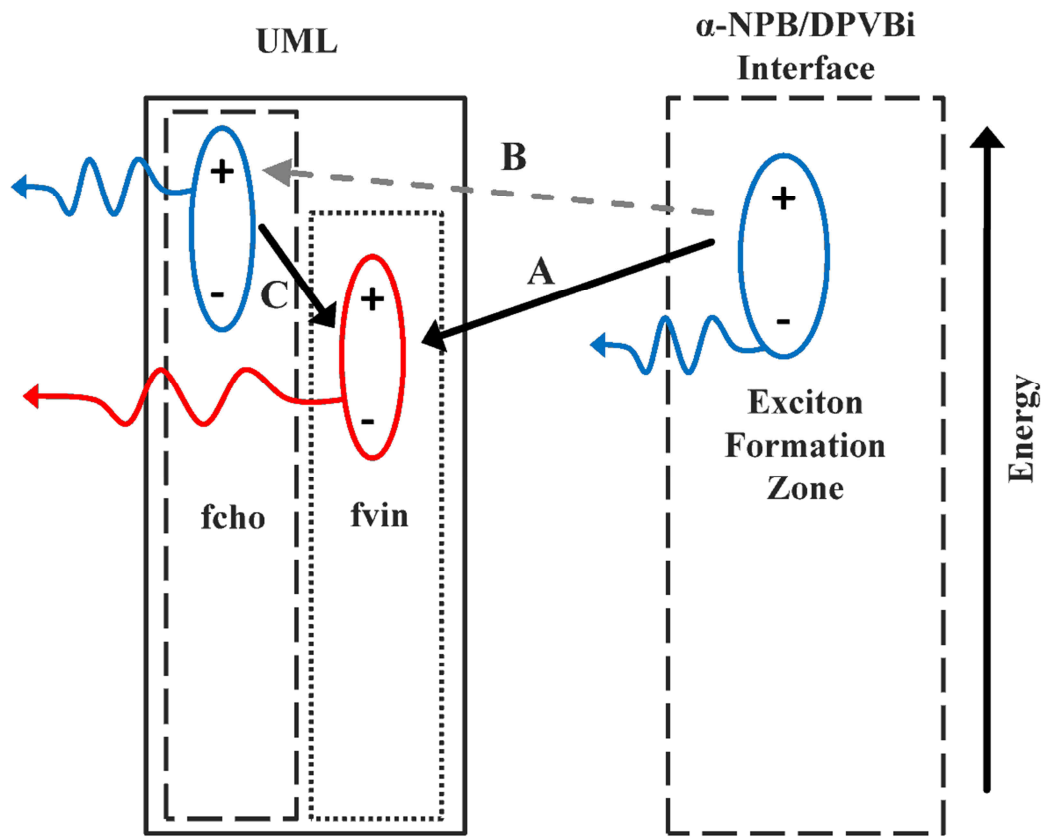

Fig.10